\documentclass[journal]{IEEEtran}
\usepackage{amsmath,amssymb,amsfonts}
\usepackage{bm}
\usepackage{cite}
\usepackage{algorithm}
\usepackage{algorithmic}
\usepackage{graphicx}
\usepackage{caption}
\usepackage{lipsum}
\usepackage{enumerate}
\usepackage{amsthm}
\usepackage{amsmath}
\usepackage{amssymb}
\usepackage{color}
\usepackage{booktabs}

\newtheorem{definition}{Definition}
\newtheorem{lemma}{Lemma}
\newtheorem{theorem}{Theorem}

\newtheorem{assumption}{Assumption}
\newtheorem{remark}{Remark}

\usepackage{bm}%专门处理数学粗体的bm宏包
\usepackage{threeparttable}
 
\ifCLASSOPTIONcompsoc
  \usepackage[caption=false,font=normalsize,labelfont=sf,textfont=sf]{subfig}
\else
  \usepackage[caption=false,font=footnotesize]{subfig}
\fi
\usepackage{tikz}
\newcommand\copyrighttext{%
  \footnotesize \textcopyright 2023 IEEE. Personal use of this material is permitted.
  Permission from IEEE must be obtained for all other uses, in any current or future
  media, including reprinting/republishing this material for advertising or promotional
  purposes, creating new collective works, for resale or redistribution to servers or
  lists, or reuse of any copyrighted component of this work in other works.}
\newcommand\copyrightnotice{%
\begin{tikzpicture}[remember picture,overlay]
\node[anchor=south,yshift=10pt] at (current page.south) {\fbox{\parbox{\dimexpr\textwidth-\fboxsep-\fboxrule\relax}{\copyrighttext}}};
\end{tikzpicture}%
}

\ifCLASSINFOpdf
\else
\fi
\begin{document}

\title{Relaxed Actor-Critic with Convergence Guarantees for Continuous-Time Optimal Control of Nonlinear Systems}

\author{ Jingliang Duan, Jie Li, Qiang Ge, Shengbo Eben Li, Monimoy Bujarbaruah, Fei Ma, Dezhao Zhang% <-this % stops a space
\thanks{Jingliang Duan and Jie Li contributed equally to this work. All correspondences should be sent to S. Li with email: lisb04@gmail.com.
}

\thanks{J. Duan is with the School of Mechanical Engineering, University of Science and Technology Beijing, Beijing, 100083, China, and also with the School of Vehicle and Mobility, Tsinghua University, Beijing, 100084, China. {\tt\small Email:duanjl15@163.com}.
}% <-this % stops a space
\thanks{J. Li, Q. Ge and S. E. Li are with the School of Vehicle and Mobility, Tsinghua University, Beijing, 100084, China. {\tt\small Email: jie-li18@mails.tsinghua.edu.cn, gq17@mails.tsinghua.edu.cn, lisb04@gmail.com}.
}% <-this % stops a space
\thanks{M. Bujarbaruah is with the Department of Mechanical Engineering, University of California Berkeley, Berkeley, CA 94720, USA. {\tt\small Email: monimoyb@berkeley.edu}.
}
\thanks{F. Ma is with the School of Mechanical Engineering, University of Science and Technology Beijing, Beijing, 100083, China. {\tt\small Email: yeke@ustb.edu.cn}.
}% <-this % stops a space
\thanks{D. Zhang is with Beijing Idriverplus Technology Co., Ltd., Beijing, 100192, China. {\tt\small Email: zhangdezhao@idriverplus.com}.
}% <-this % stops a space
}

 \maketitle
\copyrightnotice
%As a general rule, do not put math, special symbols or citations
\begin{abstract}
This paper presents the Relaxed Continuous-Time Actor-critic (RCTAC) algorithm, a method for finding the nearly optimal policy for nonlinear continuous-time (CT) systems with known dynamics and infinite horizon, such as the path-tracking control of vehicles. RCTAC has several advantages over existing adaptive dynamic programming algorithms for CT systems. It does not require the ``admissibility" of the initialized policy or the input-affine nature of controlled systems for convergence. Instead, given any initial policy, RCTAC can converge to an admissible, and subsequently nearly optimal policy for a general nonlinear system with a saturated controller. RCTAC consists of two phases: a warm-up phase and a generalized policy iteration phase. The warm-up phase minimizes the square of the Hamiltonian to achieve admissibility, while the generalized policy iteration phase relaxes the update termination conditions for faster convergence. The convergence and optimality of the algorithm are proven through Lyapunov analysis, and its effectiveness is demonstrated through simulations and real-world path-tracking tasks.
\end{abstract}


% \fi
\IEEEpeerreviewmaketitle

\section{Introduction}
Dynamic Programming (DP) offers a systematic way to solve Continuous Time (CT) infinite horizon optimal control problems with known dynamics for unconstrained linear systems. It does this by using the principle of Bellman optimality and the solution of the underlying Hamilton-Jacobi-Bellman (HJB) equation \cite{liu2017ADP}, yielding the well-known Linear Quadratic Regulator (LQR) \cite{pappas1980numerical}. In this case, the optimal control policy is an affine state feedback. However, solving an infinite horizon optimal control problem analytically becomes more difficult when the system is subject to operating constraints or has nonlinear dynamics. This is because it becomes difficult to obtain an analytical solution of the HJB equation, which is typically a nonlinear partial differential equation \cite{lewis2012OptimalControl}. This is known as the \emph{curse of dimensionality}, as the computation burden grows exponentially with the dimensionality of the system \cite{wang2009adaptive}. Traditional DP methods are therefore not applicable in these cases.

To find a nearly optimal approximation of the optimal control policy for nonlinear dynamics, Werbos proposed the Adaptive DP (ADP) method \cite{werbos1992approximate}, also known as Reinforcement Learning (RL) in the field of machine learning \cite{duan2020hierarchical,li2023RL,duan2021distributional, he2022robust,li2022probabilistic}. A distinct characteristic of ADP is that it utilizes a critic parameterized function, such as a Neural Network (NN), to approximate the value function, and an actor parameterized function to approximate the policy. The classical Value Iteration (VI) framework, which approximates the value function through one-step back-up operation, is commonly used for building ADP methods for discrete-time systems \cite{sutton2018reinforcement}. However, for CT systems, the value update law of VI requires integrating the running cost over a finite time horizon, which can lead to learning inaccuracies due to discretization error \cite{lee2013IVI}.  Most ADP methods adopt an iterative technique called Policy Iteration (PI) to find suitable approximations of both the value function and policy, which directly updates the value function by solving the corresponding Lyapunov equation without the need for discretization \cite{li2023RL}.  PI consists of two-step process: 1) policy evaluation, in which the value function moves towards its true value associated with an admissible control policy, and 2) policy improvement, in which the policy is updated to reduce the corresponding value function.

In recent years, there has been a significant amount of research on the use of ADP (or RL) techniques for the control of autonomous systems \cite{min2019deep,duan2022adaptive,fan2018robust,na2018nonlinear,sun2018backstepping}. One example of CT optimal control is the ADP algorithm proposed by Abu-Khalaf and Lewis, which seeks to find a nearly optimal constrained state-feedback controller for nonlinear systems by using a non-quadratic cost function \cite{lewis2005definition}. The value function, parameterized by a linear function of hand-crafted features, is trained by the least square method at the policy evaluation step, and the policy is expressed as an analytic function of the value function. Utilizing the same single approximator scheme, Dierks and Jagannathan developed a novel online parameter tuning law that ensures the optimality of both the value function and control policy, as well as maintaining bounded system states during the learning process \cite{dierks2010admissible}. \cite{Vamvoudakis2010OnlineAC} proposed a synchronous PI algorithm with an actor-critic framework for nonlinear CT systems without input constraints. Both the value function and policy are approximated by linear methods and tuned simultaneously online. Furthermore, Vamvoudakis introduced an event-triggered ADP algorithm that reduces computation cost by only updating the policy when a specific condition is violated \cite{vamvoudakis2014linearmethod}, and Dong \emph{et al.} expanded upon this concept for use in nonlinear systems with saturated actuators \cite{dong2017eventtrigger}. In addition, ADP methods have also been widely applied in the optimal control of incompletely known dynamic systems \cite{yang2014admissible,jiang2015global,xue2020adaptive,jiang2022bias,wang2020reinforcement} and multi-agent systems \cite{Vamvoudakis2012MultiADP, li2017off,shi2020data}.

It is worth noting that most existing ADP techniques for CT systems are valid on the basis of one or both of the following two requirements: 
\begin{itemize}
    \item  \textit{A1: Admissibility of Initial Policy}: One is the requirement for an admissible initial policy, meaning that the policy must be able to stabilize the system. This is because the infinite horizon value function can only be evaluated for stabilizing control policies. However, it can be challenging to obtain an admissible policy, particularly for complex systems.
    
    \item \textit{A2: Input-Affine Property of System}: Most ADP methods are limited to input-affine systems because these methods require that the optimal policy can be represented by the value function. This means that the minimum point of the Hamilton function can be solved analytically when the value function is given. However, this is not possible for input non-affine systems, making it difficult to directly solve the optimal policy.
\end{itemize}

Note that while there are certain methods, such as integral VI \cite{lee2013IVI} and parallel-control-based ADP methods \cite{lu2022event}, which do not require initial admissible control policies, they are only suitable for linear or input-affine systems. For example, the differentiation term of the time derivative of the Lyapunov function with respect to the control policy can be added to its updating rule to make the initial admissible policies no longer necessary \cite{yang2021decentralized,dierks2010admissible,yang2014admissible}, which requires the input-affine property. \cite{lu2020parallel} goes one step further and eliminates the restriction of the affine property by transforming the general nonlinear system into an affine system. However, this approach generally leads to some loss in policy performance. Additionally, the studies mentioned above approximate the value function or the policy using a single NN (i.e., the linear combination of predetermined hand-crafted basis functions). This means that the performance of these methods is heavily dependent on the quality of hand-crafted features, limiting their generality since it is challenging to design such features for high-dimensional nonlinear systems. Therefore, this paper aims to address these two requirements without sacrificing optimality guarantees and generality.

In this paper, we propose a relaxed continuous-time actor-critic (RCTAC) algorithm with the guarantee of convergence and optimality for solving optimal control problems of general nonlinear CT systems with known dynamics, which overcomes the limitation of the above two central requirements. Our main contributions are summarized as follows: 
\begin{enumerate}
    \item The warm-up phase of RCTAC is developed to relax the requirement of A1. It is proved that given any initial policy, when the activation function of the value network meets certain requirements, the warm-up phase can converge to an admissible policy by continuously minimizing the square of the Hamiltonian. Unlike the policy tuning rule used in \cite{yang2021decentralized,dierks2010admissible,yang2014admissible}, which also relaxes A1 but is restricted to input-affine systems and single-NN-based policies, the developed warm-up phase is applicable to general non-affine systems. Moreover, RCTAC obviates the need for designing hand-crafted basis functions by utilizing multi-layer neural networks to approximate both the actor and critic, which create mappings from the system states to control inputs and the value function, respectively.
    
    \item Different from \cite{dierks2010admissible,yang2014admissible,lee2013IVI} which require input-affine systems because the policy must be represented analytically by the value function, the policy network in the RCTAC algorithm is updated by directly minimizing the associated Hamiltonian using gradient descent methods in the generalized PI phase. This allows the RCTAC algorithm to be applied to arbitrary nonlinear dynamics with bounded and non-affine inputs, thereby relaxing the requirement of A2.  Compared to the method of transforming general nonlinear systems into affine systems by creating augmented systems \cite{lu2020parallel}, our method does not sacrifice the guarantee of theoretical optimality. 
    
    \item We introduce novel update termination conditions for the policy evaluation and improvement processes, resulting in a faster convergence speed than the corresponding PI methods. We also provide a Lyapunov analysis to prove the convergence and optimality of RCTAC.
\end{enumerate}
Throughout the paper, we provide two detailed numerical experiments to show the generality and efficacy of the proposed RCTAC algorithm, including a linear optimal control problem and a path-tracking control task for vehicles with nonlinear and non-input-affine dynamics. Besides, we demonstrate the practical application performance of our algorithm through an actual longitudinal and lateral vehicle control task.

\textbf{Organization:} Section \ref{sec.math} provides the formulation of the optimal control problem and the description of PI. In Section \ref{sec.RCTAC}, we describe the RCTAC algorithm and analyze its convergence and optimality. In Section \ref{sec.result}, we present simulation examples that show the generality and effectiveness of the RCTAC algorithm for CT system. In Section~\ref{sec.experiment}, we conduct experiments to verify the effectiveness of the method in practical applications. Section \ref{sec.conclu} concludes this paper.

\section{Mathematical Preliminaries}
\label{sec.math}

\subsection{HJB Equation}

This study considers the time-invariant system
\begin{equation}   
\label{eq.statefunction}
\dot{x}(t) = f(x(t),u(t)),
\end{equation}
where $x \in \mathbb{R}^n$ denotes the state, $u \in \mathbb{R}^m$ denotes the control input, and $f:  \mathbb{R}^n \times \mathbb{R}^m \rightarrow \mathbb{R}^n$ denotes the system dynamics. The dynamics $f (x,u)$ is assumed to be Lipschitz continuous on a compact set $\Omega$ that contains the origin. We suppose a continuous policy $u=\pi(x)$ on $\Omega$ that asymptotically stabilizes the system exists. We assume the full information of $f (x(t),u(t))$ is available. It can be represented by a nonlinear and input non-affine function or a Neural Network (NN) only if $\frac{\partial f(x,u)}{\partial u}$ is accessible. The control signal $u(t)$ can be saturated or unsaturated. The value function (cost-to-go) of policy $\pi(x)$ is calculated as
\begin{equation}   
\label{eq.costfunction}
V^{\pi}(x) = \int_{t}^{\infty} l(x(\tau),\pi(x(\tau))) {\rm{d}}\tau \Big|_{x(t)=x},
\end{equation}
where $l:  \mathbb{R}^n \times \mathbb{R}^m \rightarrow \mathbb{R}$ is positive definite, i.e., if and only if $(x,u)=(0,0)$, $l(x,u) = 0$; otherwise, $l(x,u) > 0$. 
For dynamics in \eqref{eq.statefunction} and the corresponding value function in \eqref{eq.costfunction}, we give the associated Hamilton function
\begin{equation}   
\label{eq.Hamiltonian}
H(x,u,\frac{\partial V^{\pi}(x)}{\partial x}) := l(x,u) + \frac{\partial V^{\pi}(x)}{\partial x^\top}f(x,u).
\end{equation}

\begin{definition}
\label{def.admissible}
(Admissible Policy \cite{beard1997galerkin}). A controller $\pi(x)$ is called the admissible policy, denoted by $\pi(x) \in \Psi(\Omega)$, if it is continuous on $\Omega$ with $\pi(0) = 0$, and stabilizes \eqref{eq.statefunction} on $\Omega$. 
\end{definition}
Given an admissible policy $\pi(x) \in \Psi(\Omega)$,  the differential equivalent to \eqref{eq.costfunction} on $\Omega$ is called the Lyapunov equation 
\begin{equation}   
\label{eq.Lyapunoveuqation}
H(x,\pi(x),\frac{\partial V^{\pi}(x)}{\partial x})= 0,
\end{equation}
where $V^{\pi}(0)=0$.
Therefore, given a policy $\pi(x) \in \Psi(\Omega)$, we can obtain the value function \eqref{eq.costfunction} of system \eqref{eq.statefunction} by solving the Lyapunov equation. Then the optimal control problem for Continuous Time (CT) system can be formulated as solving a policy $\pi(x) \in \Psi(\Omega)$ such that the corresponding value function is minimum.
The minimized or optimal value function $V^\star(x(t))$ defined by
\begin{equation}   
\label{eq.OCP}
V^\star(x(t)) = \min \limits_{\pi(x) \in \Psi(\Omega)}\int_{t}^{\infty} l(x(\tau),\pi(x(\tau)) {\rm{d}}\tau,
\end{equation}
satisfies the classical Hamilton-Jacobi-Bellman (HJB) equation 
\begin{equation}   
\label{eq.HJB}
\min \limits_{u} H(x,u,\frac{\partial V^\star(x)}{\partial x}) = 0, \quad V^\star(0) = 0.
\end{equation}

Meanwhile, the optimal control policy $\pi^\star(x)$ can be calculated as
\begin{equation}   
\label{eq.PMP}
\pi^\star(x) = \arg \min_u  H(x,u,\frac{\partial V^\star(x)}{\partial x}), \forall x\in\Omega. 
\end{equation}
which is the globally optimal solution to  \eqref{eq.OCP}. Inserting $V^\star(x)$ and $\pi^\star(x)$ in \eqref{eq.Lyapunoveuqation}, one can reformulate \eqref{eq.HJB} as
\begin{equation}   
\nonumber
\label{eq.HJB_2}
l(x,\pi^\star(x)) + \frac{\partial V^\star(x)}{\partial x^\top}f(x,\pi^\star(x)) = 0
\end{equation}
with  $V^\star(0)=0$. The optimal value function is shown to exist and be unique in \cite{lyashevskiy1996unique}. To obtain the optimal policy, we first need to solve the HJB equation \eqref{eq.HJB} to find the value function, and then use it to calculate the optimal policy using  \eqref{eq.PMP}. However, due to the nonlinear property of the HJB equation, it is often challenging or even impossible to find a solution.

\subsection{Policy Iteration}

Rather than attempting to solve the HJB equation directly, most Adaptive Dynamic Programming (ADP) methods adopt an iterative technique, called Policy Iteration (PI), to approximate both the value function and the policy \cite{sutton2018reinforcement}. PI consists of alternating between policy evaluation using \eqref{eq.Lyapunoveuqation} and policy improvement using \eqref{eq.PMP}. The algorithm proposed in this paper is also based on PI, as shown in the pseudocode in Algorithm \ref{alg:PI}.

\begin{algorithm}[!htb]
\caption{PI for CT optimal control}
\label{alg:PI}
\begin{algorithmic}
\STATE Initial with policy $\pi^0\in \Psi(\Omega)$
\STATE Given any small positive number $\epsilon$ and let $i=0$
\WHILE{$\max_{x\in \Omega}| V^i(x)-V^{i+1}(x)|  \ge \epsilon$}
\STATE 1. Find value function $V^i(x)$ for all $x \in \Omega$ by
\begin{equation}   
\label{eq.PE}
l(x,\pi^i(x)) + \frac{\partial V^i(x)}{\partial x^\top}f(x,\pi^i(x)) = 0, \quad  V^i(0) = 0
\end{equation} 
\STATE 2. Find new policy $\pi^{i+1}(x)$  by 
\begin{equation}  
\label{eq.PI} 
\pi^{i+1}(x) =\arg \min_u [l(x,u) + \frac{\partial V^i(x)}{\partial x^\top}f(x,u)] %\pi^{i+1}(x) =\mathop{\arg\min}\limits_{u} [l(x,u) + \frac{\partial V^i(x)}{\partial x^\top}f(x,u)] 
\end{equation}
\STATE $i=i+1$
\ENDWHILE
\end{algorithmic}
\end{algorithm}

As shown in Algorithm~\ref{alg:PI}, the first step of PI is to find an initial policy $\pi^0(x) \in \Psi(\Omega)$, because the associated value function $V^0(x)$ is finite only when the system is asymptotically stable. Algorithm \ref{alg:PI} then iteratively refines both the optimal control policy and the optimal value function. The convergence and optimality of the algorithm are proven in \cite{lewis2005definition}.

\subsection{Value Function and Policy Approximation}

In previous ADP research for CT systems, the value function $V^i(x)$ and policy $\pi^i(x)$ are usually approximated by linear methods, which requires a large number of artificially designed basis functions \cite{jiang2015global}. In recent years, NNs are favored in many fields, such as deep learning and Reinforcement Learning (RL), due to their better generality and higher fitting ability \cite{lecun2015deep}. In our work, we represent the value function and policy using multi-layer neural networks (NNs), referred to as the value NN $V(x; \omega)$ ($V_{\omega}(x)$ for short) and the policy NN  $\pi(x; \theta)$ ($\pi_{\theta}(x)$ for short), where $w$ and $\theta$ denote network weights. The two NNs in this case are used to map the original system states to the estimated value function and control inputs, respectively.

By inserting the value and policy network in \eqref{eq.Hamiltonian}, we can obtain an approximated Hamiltonian expressed in terms of the parameters $w$ and $\theta$
\begin{equation*} 
H(x,\omega,\theta)  = l(x,\pi_{\theta}(x)) + \frac{\partial V_{\omega}(x)}{\partial x^\top}f(x,\pi_{\theta}(x)).
\end{equation*}
We refer to the algorithm combining PI and multi-layer NN as approximate PI (API), which involves processes alternatively tuning each of the two networks to find nearly optimal parameters $\omega^\star$ and $\theta^\star$ satisfying $V_{\omega^\star}(x) \approx V^\star(x)$, $\pi_{ \theta^\star}(x)\approx \pi^\star(x)$.

Given any policy $\pi_{\theta} \in \Psi(\Omega)$,  the value network is tuned by solving the Lyapunov equation \eqref{eq.PE} during the policy evaluation phase of API, which is equivalent to finding parameters $w$ to minimize the following critic loss function
\begin{equation} 
\label{eq.loss_c}
L_c(\omega, \theta) = \mathbb{E}_{x\sim d_x}\big[H(x,\omega,\theta)^2\big],
\end{equation}
where $d_x$ represents the state distribution over $\Omega$. It is important to note that $d_x$ can be any distribution that satisfies the requirement of a positive probability density for all $x \in \Omega$, such as the uniform distribution. The condition $V(x; \omega) \equiv 0$ can be easily guaranteed by selecting proper activation function $\sigma_V(\cdot)$ for the value network. Based on \eqref{eq.PI}, the policy improvement process is carried out by tuning the policy network to minimize the expected Hamiltonian in each state, which is also called actor loss function here
\begin{equation} 
\label{eq.loss_a}
L_a(\omega, \theta) = \mathbb{E}_{x\sim d_x}\big[H(x,\omega,\theta)\big].
\end{equation}
The benefit of updating the policy network by minimizing $L_a(\omega, \theta)$ is that the tuning rule is applicable to any nonlinear dynamics with non-affine and constrained inputs. This relaxes the requirement of A2 (from Introduction).

Since the state $x$ is continuous, it is difficult to directly compute the expectation in \eqref{eq.loss_c} and \eqref{eq.loss_a}. In practice, these two loss functions can be estimated by the sample average. We can directly utilize existing NN optimization methods to adjust the parameters of value and policy NNs, such as Stochastic Gradient Descent (SGD). The value network and policy network usually require multiple updating iterations to make \eqref{eq.PE} and \eqref{eq.PI} hold respectively. Therefore, compared with PI algorithm, two inner updating loops would be introduced to update the value and policy NNs until convergence. Taking SGD as an example, we give the pseudocode of API in Algorithm~\ref{alg:API}. 
\begin{algorithm}[!htb]
\caption{API for CT optimal control}
\label{alg:API}
\begin{algorithmic}
\STATE Initial with $\theta^0$ such that $\pi_{\theta^0}(x) \in \Psi(\Omega)$ and arbitrary $\omega^{0}$
\STATE Choose the appropriate learning rates $\alpha_{\omega}$ and $\alpha_{\theta}$
\STATE Given any small positive number $\epsilon$ and set $i=0$
\WHILE{$\max_{x\in \Omega}| V_{\omega^{i}}(x)-V_{\omega^{i-1}}(x) | \ge \epsilon$}
\STATE 1. Estimate $V_{\omega^{i+1}}(x)$ using $\pi_{\theta^i}(x)$

\setlength{\leftskip}{2em}
\STATE $\omega^{i+1}= \omega^{i}$
\REPEAT
\STATE 
\begin{equation}   
\label{eq.SPE}
\omega^{i+1} = \omega^{i+1} - \alpha_{\omega} \frac{{\rm{d}} L_c(\omega^{i+1}, \theta^i)}{{\rm{d}}\omega^{i+1}} 
\end{equation}
\UNTIL{$L_c(\omega^{i+1}, \theta^i)  \le \epsilon$}

\setlength{\leftskip}{0em}
\STATE 2. Find improved policy $\pi_{\theta^{i+1}}(x)$ using $V_{\omega^{i+1}}(x)$

\setlength{\leftskip}{2em}
$\theta^{i+1}= \theta^{i}$\;
\REPEAT
\STATE 
\begin{equation}  
\nonumber
L_{a,\rm{old}}=L_a(\omega^{i+1}, \theta^{i+1})
\end{equation}
\begin{equation}  
\label{eq.SPI}
\theta^{i+1} = \theta^{i+1} - \alpha_{\theta} \frac{{\rm{d}}L_a(\omega^{i+1}, \theta^{i+1})}{{\rm{d}}\theta^{i+1}} 
\end{equation}
\UNTIL{$|L_a(\omega^{i+1}, \theta^{i+1})-L_{a,\rm{old}}|  \le \epsilon$}
\STATE $i=i+1$
\setlength{\leftskip}{0em}
\ENDWHILE
\end{algorithmic}
\end{algorithm}

\section{Relaxed Continuous-Time Actor-Critic}
\label{sec.RCTAC}
% \subsection{Motivation}
Algorithm~\ref{alg:API} alternately update the value and policy network by minimizing \eqref{eq.loss_c} and \eqref{eq.loss_a}, respectively. Note that while one NN is being adjusted, the other remains constant. Besides, each NN generally needs multiple iterations to satisfy the terminal condition, which is referred to as the protracted iterative computation problem \cite{sutton2018reinforcement}. This problem usually leads to the admissibility requirement because the initial policy network needs to satisfy that $\pi(x;\theta^0) \in \Psi(\Omega)$ to have a finite and converged value function $V(x;\omega^{1})$. Many previous studies used trials and errors process to obtain feasible initial weights for the policy network to guarantee the stability of the system \cite{liu2017ADP,vamvoudakis2014linearmethod}. However, this method usually takes a lot of time, especially for complex systems. On the other hand, the protracted problem often results in slower learning \cite{sutton2018reinforcement}.

\subsection{Description of the RCTAC Algorithm}

Inspired by the idea of generalized PI framework, which is widely utilized in discrete-time dynamic RL problems \cite{sutton2018reinforcement}, we present the relaxed continuous-time actor-critic (RCTAC) algorithm for CT systems to relax the requirement A1 (from Introduction) and improve the learning speed by truncating the inner loops of Algorithm~\ref{alg:API} without losing the convergence guarantees. The pseudocode of RCTAC algorithm is shown in Algorithm~\ref{alg:RCTAC}.

\begin{algorithm}[!htb]
\caption{RCTAC algorithm}
\label{alg:RCTAC}
\begin{algorithmic}
\STATE Initialize arbitrary $\theta^0$ and $\omega^{0}$
\STATE Choose the appropriate learning rates $\alpha$, $\alpha_{\omega}$ and $\alpha_{\theta}$
\STATE Given any small positive number $\epsilon$ and set $i=0$
\STATE Phase 1: Warm-up

\setlength{\leftskip}{1em}
\WHILE{$\max_{x\in \Omega}H(x,\omega^{i},\theta^{i})\ge 0$}
\setlength{\leftskip}{1em}
\STATE Update $\omega$ and $\theta$ using:
\begin{equation}
\label{eq.warmup}
\{\omega^{i+1},\theta^{i+1}\} = \{\omega^{i},\theta^{i}\} - \alpha \frac{{\rm{d}} L_c(\omega^{i}, \theta^i)}{{\rm{d}}\{\omega^{i},\theta^{i}\}}\; 
\end{equation}
\STATE $i=i+1$
\ENDWHILE

\setlength{\leftskip}{0em}
\STATE Phase 2: PI with relaxed termination conditions

\setlength{\leftskip}{1em}
\WHILE{$\max_{x\in \Omega}| V_{\omega^{i}}(x)-V_{\omega^{i-1}}(x)|  \ge \epsilon$}
\setlength{\leftskip}{1em}
\STATE 1. Estimate $V_{\omega^{i+1}}(x)$ using $\pi_{\theta^i}(x)$

\setlength{\leftskip}{3em}
\STATE $\omega^{i+1}= \omega^{i}$
\REPEAT
\setlength{\leftskip}{3em}
\STATE Update $\omega^{i+1}$ using \eqref{eq.SPE}
\UNTIL{$H(x,\omega^{i},\theta^{i}) \le H(x,\omega^{i+1},\theta^{i}) \le 0, \forall x\in \Omega$}

\setlength{\leftskip}{1em}
\STATE 2. Find improved policy $\pi_{\theta^{i+1}}(x)$ using $V_{\omega^{i+1}}(x)$ 

\setlength{\leftskip}{3em}
\STATE $\theta^{i+1}= \theta^{i}$
\REPEAT
\setlength{\leftskip}{3em}
\STATE Update $\theta^{i+1}$ using \eqref{eq.SPI}
\UNTIL{$\max_{x \in \Omega}H(x,\omega^{i+1},\theta^{i+1})\le 0$}

\setlength{\leftskip}{1em}
\STATE $i=i+1$

\setlength{\leftskip}{0em}
\ENDWHILE
\end{algorithmic}
\end{algorithm}

\subsection{Convergence and Optimality Analysis}

The solution to \eqref{eq.PE} may be non-smooth for general nonlinear and input non-affine systems. In keeping with other research in the literature \cite{Vamvoudakis2010OnlineAC}, we make the following assumptions. 

\begin{assumption}
\label{assum.smooth}
If $\pi(x) \in \Psi(\Omega)$, its corresponding value function is smooth, i.e. $V^{\pi}(x) \in C^1 (\Omega)$ (cf. \cite{lewis2005definition,Vamvoudakis2010OnlineAC}).
\end{assumption}

Multiple theoretical analyses and experimental studies have demonstrated that optimization algorithms like SGD are capable of locating the global minimum of the training objective for multi-layer NNs in polynomial time, provided that the NN is over-parameterized (with a sufficiently large number of hidden neurons) \cite{allen2018convergence,du2018gradient}. Based on this fact, our second assumption is:

\begin{assumption}
\label{assum.optimal}
Optimization algorithms like SGD can locate the global minimum of the critic loss function \eqref{eq.loss_c} and the actor loss function \eqref{eq.loss_a} for over-parameterized NNs in polynomial time.
\end{assumption}

In the following section, we will demonstrate that the nearly optimal value function and policy can be obtained using Algorithm \ref{alg:RCTAC}. To do so, we will first introduce some necessary lemmas.

\begin{lemma} 
\label{lemma.approximation}
(Universal Approximation \cite{Hornik1990Universal}). For any continuous function $F(x)$ defined on a compact set $\Omega$, it is possible to construct a feedforward NN with one hidden layer that uniformly approximates $F(x)$ with arbitrarily small error $\epsilon \in \mathbb{R}^{+}$.
\end{lemma}
Lemma \ref{lemma.approximation} allows us to ignore the NN approximation errors when proving the convergence of Algorithm~\ref{alg:RCTAC}.
\begin{lemma}
\label{lemma.optimality}
Consider the CT dynamic optimal control problem for \eqref{eq.statefunction} and \eqref{eq.costfunction}. Suppose the solution ($V^{\pi}(x) \in C^1: \mathbb{R}^n \rightarrow \mathbb{R} $ ) of the HJB equation \eqref{eq.HJB} is smooth and positive definite. The control policy $\pi(x)$ is given by \eqref{eq.PMP}. Then we have that $V^{\pi}(x) = V^\star(x)$ and $\pi(x) = \pi^\star(x) $ (cf. \cite{lewis2012OptimalControl}). %P450
\end{lemma} 
The next lemma shows how Algorithm~\ref{alg:RCTAC} can be leveraged to obtain an admissible policy $\pi(x; \theta) \in \Psi(\Omega)$ given any initial policy $\pi(x; \theta^0)$.

\begin{lemma}
\label{lemma.admissible}
Consider the CT dynamic optimal control problem for \eqref{eq.statefunction} and \eqref{eq.costfunction}. The value function (cost-to-go) $V_{\omega}$ and policy $\pi_{\theta}$ are represented by over-parameterized NNs. The parameters $w$ and $\theta$ are initialized randomly, i.e., the initial policy $\pi(x; \theta^0)$ can be inadmissible. These two NNs are updated with Algorithm~\ref{alg:RCTAC}. Let Assumption \ref{assum.smooth} and \ref{assum.optimal} hold, and suppose all the hyper-parameters (such as $\alpha$, $\alpha_{w}$ and $\alpha_{\theta}$) and NN optimization method are properly selected. The NN approximation errors are ignored according to Lemma \ref{lemma.approximation}. Suppose all the activation functions $\sigma_V(\cdot)$ and biases $b_V$ of the value network $V(x; \omega)$ are set to $\sigma_V(0)=0$ and $b_V(\cdot) \equiv 0$. At the same time, the output layer activation function $\sigma_{V_\textnormal{out}}$ also needs to satisfy $\sigma_{V_\textnormal{out}}(\cdot) \ge 0$. We have that: $\exists N_a \in \mathbb{Z}^+$, if the iteration index $i\ge N_a$, then $\pi(x;\theta^{i})\in \Psi(x)$ for the system \eqref{eq.statefunction} on $\Omega$. 
\end{lemma}

\begin{proof}
According to \eqref{eq.Lyapunoveuqation} and Lemma~\ref{lemma.approximation}, there $\exists (\omega^{\dagger}, \theta^{\dagger})$, such that $\pi(x;\theta^{\dagger})\in \Psi(\Omega)$ and $H(x,\omega^{\dagger},\theta^{\dagger})= 0$ for all $x \in \Omega$, which implies that 
\begin{equation*}
% \label{eq.minH}
\min_{\omega,\theta}L_c(\omega,\theta) = \min_{\omega,\theta} \mathbb{E}_{x\sim d_x}\Big[H(x,\omega,\theta)^2\Big]=0.
\end{equation*}
% Similarly, we can also get that $\exists (\omega^{\ddagger}, \theta^{\ddagger})$, such that 
% \begin{equation}
% \nonumber
% H(x,\omega^{\ddagger},\theta^{\ddagger})\le 0, \quad \forall x \in \Omega.
% \end{equation} 

Since Algorithm~\ref{alg:RCTAC} updates $\omega$ and $\theta$ using \eqref{eq.warmup} to continuously minimize $L_c(\omega,\theta)$ in Phase 1, according to Assumption \ref{assum.optimal}, the global minima of $L_c(\omega,\theta)$ can be obtained in polynomial time. Before the global minima is found, there exists $N_a \in \mathbb{Z}^+$, such that
\begin{equation}
\label{eq.H<0}
H(x,\omega^{N_a},\theta^{N_a}) \le 0, \quad \forall x\in\Omega.
\end{equation}
Taking the time derivative of $V(x; \omega)$, we can derive that
\begin{equation}   
\label{eq.dv/dt}
\begin{split}
\frac{{\rm{d}}V(x; \omega)}{{\rm{d}}t}  &=  \frac{\partial V(x; \omega)}{\partial x^\top} f(x,\pi(x;\theta)),  \\
&= H(x,\omega,\theta) - l(x, \pi(x;\theta)).
\end{split}
\end{equation} 
Using \eqref{eq.H<0} and \eqref{eq.dv/dt}, one has
\begin{equation}
\nonumber
\frac{{\rm{d}}V(x; \omega^{N_a})}{{\rm{d}}t} \le  - l(x, \pi(x;\theta^{N_a})), \quad \forall x \in \Omega.
\end{equation} 
As the running cost $l(x, \pi(x;\theta))$ is positive definite, it follows
\begin{equation}
\label{eq.lyapunov}
\frac{{\rm{d}}V_{\omega^{N_a}}(x)}{{\rm{d}}t} < 0, \quad \forall x \in \Omega \backslash \{0\}.
\end{equation} 
Since $\sigma_V(0)=0$, $b_V(\cdot) \equiv 0$ and $\sigma_{V_\textnormal{out}}(\cdot) \ge 0$, we have
\begin{equation}
\left\{
\begin{aligned}
&V(x;\omega)  = 0  \text{\ if\ }  x = 0 \ \&\  \forall \omega,\\
&V(x;\omega)  \ge 0 \text{\ if\ } \forall x \in \Omega \backslash \{0\} \ \&\  \forall \omega.
\end{aligned}
\right.
\label{eq.Vge0}
\end{equation}
From \eqref{eq.lyapunov} and \eqref{eq.Vge0}, we have
\begin{equation}
\label{eq.Vneq0}
V(x; \omega^{N_a}) > \min_{z \in \Omega}V(z; \omega^{N_a}) =  0, \quad  \forall x \in \Omega \backslash \{0\}.
\end{equation} 
From \eqref{eq.Vge0} and \eqref{eq.Vneq0}, we infer that the value function $V(x; \omega^{N_a})$ is positive definite. Then, according to \eqref{eq.lyapunov}, $V(x;\omega^{N_a})$ is a Lyapunov function for the closed-loop dynamics obtained from \eqref{eq.statefunction} when policy $\pi(x;\theta^{N_a})$ is used. Therefore, the policy $\pi(x;\theta^{N_a}) \in \Psi(\Omega)$ for the system \eqref{eq.statefunction} on $\Omega$ \cite{Lyapunov1993stability}, that is, it is a stabilizing admissible policy. 

At this point, Algorithm~\ref{alg:RCTAC} enters Phase 2. According to \eqref{eq.Lyapunoveuqation}, one has
\begin{equation}
\nonumber
\min_{\omega}L_c(\omega,\theta^{N_a}) =\min_{\omega} \mathbb{E}_{x\sim d_x}\Big[H(x,\omega,\theta^{N_a})^2\Big]= 0.
\end{equation} 
So, from Assumption \ref{assum.optimal} and Lemma \ref{lemma.approximation}, we can always find $\omega^{N_a+1}$ by continuously applying \eqref{eq.SPE}, such that
\begin{equation}
\nonumber
H(x,\omega^{N_a},\theta^{N_a}) \le H(x,\omega^{N_a+1},\theta^{N_a}) \le 0, \quad \forall x \in \Omega.
\end{equation} 
Again, from Lemma~\ref{lemma.approximation}, there always $\exists \theta^{\#}$, such that 
\begin{equation}
\nonumber
H(x,\omega^{N_a+1},\theta^{\#}) =  \min_{\theta}H(x,\omega^{N_a+1},\theta), \quad \forall x \in \Omega.
\end{equation}
Since
\begin{equation}
\nonumber
\min_{\theta}L_a(\omega^{N_a+1},\theta) \ge  \mathbb{E}_{x\sim d_x}\Big[\min_{\theta}H(x,\omega^{N_a+1},\theta)\Big],
\end{equation}
we have
\begin{equation}
\nonumber
\begin{aligned}
L_a(\omega^{N_a+1},\theta^{\#}) &= \min_{\theta}L_a(\omega^{N_a+1},\theta)\\
&=\mathbb{E}_{x\sim d_x}\Big[\min_{\theta}H(x,\omega^{N_a+1},\theta)\Big].
\end{aligned}
\end{equation}
Utilizing the fact that the global minima of $L_a(\omega^{N_a+1},\theta)$ can be obtained, Hamiltonian $H(x,\omega^{N_a+1},\theta)$ can be taken to global minimum for all $x \in \Omega$ by minimizing over $\theta$. Then, we can also find $\theta^{N_a+1}$ through \eqref{eq.SPI}, such that
\begin{equation}
\nonumber
H(x,\omega^{N_a+1},\theta^{N_a+1}) \le H(x,\omega^{N_a+1},\theta^{N_a}) \le 0, \quad \forall x \in \Omega.
\end{equation} 
This implies that like the case with $V(x; \omega^{N_a})$, $V(x; \omega^{N_a+1})$ is also a Lyapunov function. So, $\pi(x;\theta^{N_a+1}) \in \Psi(\Omega)$. Applying this for all subsequent time steps, $V(x; \omega^{i})$ is a Lyapunov function for $\forall i \ge N_a$, and it is obvious that
\begin{equation}
\label{eq.Hincrease}
H(x,\omega^{i},\theta^{i}) \le H(x,\omega^{i+1},\theta^{i}) \le 0, \quad \forall i \ge N_a \ \&\  \forall x \in \Omega,
\end{equation} 
and
\begin{equation}
\label{eq.adimiisble_policy}
\pi(x;\theta^{i}) \in \Psi(\Omega), \quad \forall i \ge N_a.
\end{equation} 
Thus, we have proved that starting from any initial policy, the RCTAC algorithm in Algorithm~3 converges to an admissible policy. As claimed previously, this relaxes the requirement A1, which is typical for most other ADP algorithms. 
\end{proof}

We now present our main result, which shows that the value NN $V(x;\omega)$ and policy NN $\pi(x;\theta)$ converge to optimum uniformly by applying Algorithm~\ref{alg:RCTAC}. 

\begin{definition}
\label{def.uniform convergence}
(Uniform Convergence). A function sequence $\{f_n\}$ converges uniformly to $f$ on a set $\mathbb{K}$ if $\forall \epsilon>0$, $\exists N(\epsilon)\in \mathbb{Z}^{+}$, s.t. $\forall n > N$, $\sup_{x\in\mathbb{K}}\left| f_n(x)-f(x) \right | < \epsilon$. 
\end{definition}

\begin{theorem}
\label{theo.converge}
For arbitrary initial value network $V(x; \omega^{0})$ and policy network $\pi(x; \theta^0)$, where all the activation functions $\sigma_V(\cdot)$ and biases $b_V$ of the value network are set to $\sigma_V(0)=0$ and $b_V(\cdot) \equiv 0$, and the output layer activation function $\sigma_{V_\textnormal{out}}$ also satisfies $\sigma_{V_\textnormal{out}}(\cdot) \ge 0$, if these two NNs are tunned with Algorithm~\ref{alg:RCTAC}, $V_{\omega^i}(x) \rightarrow V^\star(x)$, $\pi_{\theta^i}(x) \rightarrow \pi^\star(x)$ uniformly on $\Omega$ as $i \rightarrow \infty$. 
\end{theorem}

\begin{proof}
From Lemma \ref{lemma.admissible}, it can be shown by induction that the policy $\pi(x;\theta^{i}) \in \Psi(\Omega)$ for system \eqref{eq.statefunction} on $\Omega$ when $i \ge N_a$. Furthermore, according to \eqref{eq.dv/dt} and \eqref{eq.Hincrease}, given policy $\pi(x;\theta^i)$, we get
\begin{equation}
\label{eq.decreasedV/dt}
\frac{{\rm{d}}V(x; \omega^{i})}{{\rm{d}}t} \le \frac{{\rm{d}}V(x; \omega^{i+1})}{{\rm{d}}t} \le 0 , \quad \forall x \in \Omega \ \&\  i \ge N_a.
\end{equation} 
From Newton-Leibniz formula, 
\begin{equation}
\label{eq.Newton-Leibniz}
V_{\omega}(x(t)) = V_{\omega}(x(\infty)) - \int_{t}^{\infty} \frac{{\rm{d}}V_{ \omega}(x(\tau))}{{\rm{d}}\tau} {\rm{d}}\tau.
\end{equation} 
According to \eqref{eq.Vge0} and \eqref{eq.adimiisble_policy}, 
\begin{equation}
\label{eq.zeropoint}
V_{\omega}(x(\infty)) = V_{\omega}(0) = 0, \quad i \ge N_a \ \&\  \forall \omega.
\end{equation} 
So, from \eqref{eq.Vge0}, \eqref{eq.decreasedV/dt},  \eqref{eq.Newton-Leibniz} and \eqref{eq.zeropoint}, we have
\begin{equation}
\label{eq.Vdecrease}
0\le V(x;\omega^{i+1}) \le  V(x;\omega^{i}), \quad \forall x \in \Omega \ \&\  i \ge N_a.
\end{equation} 
As such, $V(x;\omega^i)$ is pointwise convergent as $i$ goes to $\infty$. One can write $\lim_{i \rightarrow \infty}V(x;\omega^{i})=V(x;\omega^{\infty})$. Then, the compactness of $\Omega$ directly leads to uniform convergence by Dini’s theorem \cite{bartle2011Dinistheorem}. 

Next, from Definition \ref{def.uniform convergence}, it is not hard to show that
\begin{equation}
\nonumber
\lim_{i \rightarrow \infty}\sup_{x \in \Omega }  | V(x;\omega^{i})  - V(x;\omega^{i+1})|=0.
\end{equation}
Furthermore, since
\begin{equation}
\nonumber
\begin{aligned}
V(x;\omega^{i}) &- V(x;\omega^{i+1}) \\
&= \int_{t}^{\infty} \frac{{\rm{d}}(V(x(\tau); \omega^{i+1})-V(x(\tau); \omega^i))}{{\rm{d}}\tau}{\rm{d}}\tau,\\
& = \int_{t}^{\infty}\Big[H(x(\tau),\omega^{i+1},\theta^{i}) - H(x(\tau),\omega^{i},\theta^{i})\Big]{\rm{d}}\tau,
\end{aligned}
\end{equation} 
we have
\begin{equation}
\label{eq.Hconverge}
\lim_{i \rightarrow \infty}\sup_{x \in \Omega }  | H(x,\omega^{i+1},\theta^{i}) - H(x,\omega^{i},\theta^{i}) |=0.
\end{equation} 
From Lemma \ref{lemma.approximation}, \eqref{eq.Lyapunoveuqation} and \eqref{eq.adimiisble_policy}, 
\begin{equation}
\nonumber
\min_{\omega}L_c(\omega, \theta^i)= \min_{\omega} \mathbb{E}_{x\sim d_x}\Big[H(x,\omega,\theta^i)^2\Big]=0, \quad \forall i\ge N_a.
\end{equation} 
Since $\omega^{i+1}$ is obtained by minimizing $L_c(\omega^i, \theta^i)$ using \eqref{eq.SPE}, according to \eqref{eq.Hconverge}, it is true that 
\begin{equation}
\label{eq.LE=0}
\lim_{i \rightarrow \infty}H(x, \omega^{i}, \theta^{i})=\lim_{i \rightarrow \infty}H(x, \omega^{i+1}, \theta^{i})=0, \quad \forall x \in \Omega.
\end{equation} 
Therefore, $V(x;\omega^{\infty})$ is  the solution of the Lyapunov equation \eqref{eq.Lyapunoveuqation} when a policy $\pi(x; \theta^{\infty})$ is given, and it leads to that 
\begin{equation}
\nonumber
\label{eq.Vinfinity}
V_{\omega^{\infty}}(x) = V^{\pi_{\theta^{\infty}}}(x).
\end{equation} 
The policy $\pi(x;\theta^i) \in \Psi(\Omega)$ for $i \ge N_a$, so the generated state trajectory is unique due to the locally Lipschitz continuity of the dynamics \cite{lewis2005definition}. Since $V(x;\omega^i)$ converges uniformly to $V(x;\omega^{\infty})$, its obvious that the system trajectory converges for all $x \in \Omega$. Therefore, $\pi(x; \theta^i)$ also converges uniformly to $\pi(x; \theta^{\infty})$ on $\Omega$. Since $\theta^{i+1}$ is obtained by minimizing $L_a(\omega^{i+1}, \theta^i)$ using \eqref{eq.SPI}, it is obvious from \eqref{eq.LE=0} that
\begin{equation}
\lim_{i \rightarrow \infty}L_a(\omega^{i+1},\theta^{i+1})=\lim_{i \rightarrow \infty}L_a(\omega^{i+1},\theta^{i})=0,
\end{equation}
which implies that 
\begin{equation}
\label{eq.P_optimal}
\lim_{i \rightarrow \infty}\min_{\theta}H(x, \omega^{i}, \theta) = 0, 
\quad \forall x \in \Omega. 
\end{equation}
From \eqref{eq.LE=0}, \eqref{eq.P_optimal}, and Lemma \ref{lemma.optimality}, one has $\lim_{i \rightarrow \infty}V(x;\omega^{i}) = V^\star(x)$ and $\lim_{i \rightarrow \infty} \pi(x;\theta^{i}) =  \pi^\star(x)$. Therefore, one can conclude that $V_{\omega^i}(x)$ goes to $V^\star(x)$ and $\pi_{\theta^i}(x)$ goes to $\pi^\star(x)$ uniformly on $\Omega$ as $i\rightarrow \infty$, which means the global optimal solution is obtained.
\end{proof}
This demonstrates that the RCTAC algorithm uniformly converges to the optimal value function $V^\star(x)$ and the optimal policy $\pi^\star(x)$. 

\begin{remark}
The warm-up phase of RCTAC is developed to find the initial admissible policy, whose purpose is akin to that of the relaxing initial condition of discrete-time systems \cite{Liu2015Generalized}. The difference is that the latter needs to repeatedly select an initial positive value function until the initial admissible policy can be constructed. It is proved that under mild restrictions of the activation functions of the value function, given any initial policy, the warm-up phase can converge to an admissible policy by continuously minimizing the square of the Hamiltonian.
\end{remark}

\begin{remark}
According to the implementation process \eqref{eq.SPE} and \eqref{eq.SPI} of the proposed RCTAC algorithm, the dynamic $f(x,u)$ can be any analytic function such that $\frac{\partial f(x,u)}{\partial u}$ is available. As a result, RCTAC can be applied to any nonlinear system with non-affine and bounded control inputs, unlike \cite{dierks2010admissible,yang2014admissible,lee2013IVI} which are only applicable to systems with affine control inputs due to the requirement for an analytical expression of the control policy using the value function. Control constraints can be easily incorporated by using a saturated activation function (such as $\rm tanh(\cdot)$) at the output of the policy network.
\end{remark}

\begin{figure}[!htb]
\centering
\includegraphics[width=.95\linewidth]{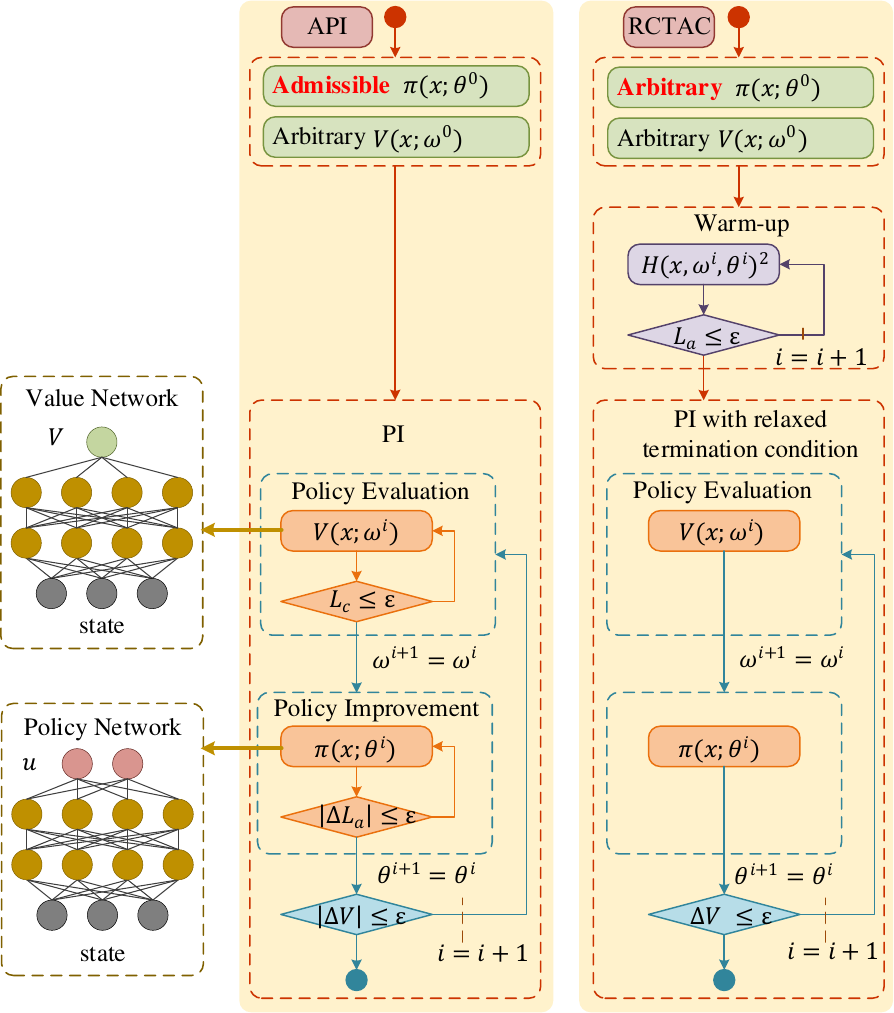}
\caption{API and RCTAC algorithm framework diagram.}
\label{fig:framework}
\end{figure}

\begin{remark}
Since the state $x$  is continuous, it is intractable to check the value of $H(x,\omega,\theta)$ for every $x \in \Omega$. Therefore, in practical applications, we usually use the expected value of $H(x,\omega,\theta)$ to judge whether each termination condition in Algorithm~\ref{alg:RCTAC} is satisfied. So, the RCTAC algorithm can also be formulated as Algorithm~\ref{alg:RCTAC2}. Fig. \ref{fig:framework} shows the frameworks of API Algorithm~\ref{alg:API} and RCTAC Algorithm~\ref{alg:RCTAC2}.
\end{remark}

\begin{algorithm}[!htb]
\caption{RCTAC algorithm: Tractable Relaxation}
\label{alg:RCTAC2}
\begin{algorithmic}
\STATE Initialize arbitrary $\theta^0$ and $\omega^{0}$
\STATE Choose the appropriate learning rates $\alpha$, $\alpha_{\omega}$ and $\alpha_{\theta}$
\STATE Given any small positive number $\epsilon$ and set $i=0$
\STATE Phase 1: Warm-up

\setlength{\leftskip}{1em}
\WHILE{$L_a(\omega^{i},\theta^{i})\ge \epsilon$}
\setlength{\leftskip}{2em}
\STATE Update $\omega$ and $\theta$ using  \eqref{eq.warmup}
\STATE $i=i+1$
\ENDWHILE

\setlength{\leftskip}{0em}
\STATE Phase 2: PI with relaxed termination conditions

\setlength{\leftskip}{1em}
\WHILE{$\mathbb{E}_{x\sim d_x}| V_{\omega^{i}}(x)-V_{\omega^{i-1}}(x) |  \ge \epsilon$}
\setlength{\leftskip}{2em}
\STATE Update $w^{i+1}$ using \eqref{eq.SPE}
\setlength{\leftskip}{2em}
\STATE Update $\theta^{i+1}$ using \eqref{eq.SPI}
\STATE $i=i+1$
\ENDWHILE

\setlength{\leftskip}{0em}
\end{algorithmic}
\end{algorithm}

It is worth noting that for Algorithm \ref{alg:RCTAC2}, sometimes skipping Phase 1 and directly using Phase 2 can also obtain a nearly optimal policy. This is because even if $L_a(\omega^i,\theta^i)\ge \epsilon$, Phase 2 would continuously make $L_a(\omega^i,\theta^i)$ approach a non-positive number by alternately using \eqref{eq.SPE} to optimize $V(x;\omega)$ and using \eqref{eq.SPI} to optimize $\pi(x;\theta)$. Note that the Phase 2 of Algorithm \ref{alg:RCTAC2} is not a CT version of the Value Iteration (VI) method. As Lee \emph{et al.} pointed out in \cite{lee2013IVI}, all VI methods for CT optimal control can be deemed as a variant of integral VI. The integral ADP methods, such as integral VI, integral PI and integral generalized PI, iteratively perform policy evaluation and policy improvement updates relying on the running cost during a finite time interval \cite{lee2013IVI}, which is clearly different from Algorithm \ref{alg:RCTAC} and Algorithm \ref{alg:RCTAC2}. Besides, these integral ADP methods are usually subject to input-affine systems since these methods require that the optimal policy can be directly represented by the value function, which means that the minimum point of the Hamilton function could be solved analytically when the value function is given. This manner is difficult to extend to input non-affine systems.

\begin{remark}
\label{remark.Hnegzero}
In the previous analysis, the utility function $l(x,u)$ is limited to positive definite functions, i.e., the equilibrium state (denoted by $x_e$) of the system must be $x_e = 0$. If we take $x - x_e$ as the input of value network $V(x; \omega)$, the RCTAC Algorithm~\ref{alg:RCTAC2} can be extended to problems with non-zero $x_e$, where $l(x, u)=0$ only if $x = x_e \ \& \ u = 0$. The corresponding convergence and optimality analysis is similar to the problem of $x_e=0$.
\end{remark}

\begin{remark}
\label{remark.netarchiteture}
According to Lemma \ref{lemma.admissible}, all activation functions $\sigma_V$ and biases $b_V$ of $V(x; \omega)$ are set to $\sigma_V(0)=0$ and $b_V(\cdot) \equiv 0$ to ensure $V(x_e; \omega) \equiv 0$. To remove these restrictions for value networks, we propose another effective method that drives $V(x_e; \omega)$ to gradually approach 0 by adding an equilibrium term to the critic loss function \eqref{eq.loss_c}
\begin{equation}
\nonumber
\label{eq.modifylv}
{L_c}'(\omega, \theta) = \mathbb{E}_{x\sim d_x} \big[H(x,\omega,\theta)^2 \big]+ \eta V(x_e; \omega)^2,
\end{equation}  
where $\eta$ balances the importance of the Hamiltonian term and the equilibrium term.
\end{remark}

\section{Results}
\label{sec.result}
To verify the proposed RCTAC Algorithm~\ref{alg:RCTAC2}, we offer two numerical examples, one with linear dynamics, and the other one with a nonlinear input non-affine system (the vehicle path-tracking control). We apply Algorithm~\ref{alg:RCTAC2} and Algorithm~\ref{alg:API} to find the optimal solutions for these two systems. Results indicate that our algorithm outperforms Algorithm~\ref{alg:API} in both cases.

\subsection{ Example I: Linear Time Invariant System}
\subsubsection{Description}
Consider the CT aircraft plant control problem used in \cite{stevens2015aircraft,Vamvoudakis2010OnlineAC,vamvoudakis2014linearmethod}, which can be formulated as 
\begin{equation}
\nonumber
\min_u \quad \int_{0}^{\infty} (x^\top Qx+u^\top Ru){\rm{d}}t
\end{equation} 
\begin{equation}
\nonumber
\rm{s.t.} \quad \dot{x} = 
\begin{bmatrix}
  -1.01887 & 0.90506 & -0.00215 \\
  0.82225 & -1.07741 & -0.17555 \\
  0 & 0 & -1
\end{bmatrix} 
x+
\begin{bmatrix}
  0 \\
  0 \\
  1 
\end{bmatrix} 
u,
\end{equation} 
where $Q$ and $R$ are identity matrices of proper dimensions. For this linear case, the optimal analytic policy $\pi^\star(x)=0.1352x_1+0.1501x_2-0.4329x_3$ and the optimal value function $V^\star(x)=x^\top Px$ can be easily found by solving the algebraic Riccati equation, where 
\begin{equation}
\label{eq.P_matrix}
\nonumber
P=
\left[
\begin{matrix}
    1.4245  &  1.1682  &  -0.1352 \\
    1.1682  &  1.4349  &  -0.1501 \\
   -0.1352  &  -0.1501 &   0.4329
 \end{matrix}
 \right].
\end{equation} 
\subsubsection{Algorithm Details}
This system is very special, in particular, if the weights of the policy NN are randomly initialized around 0, which is a very common initialization method, then the initialized policy $\pi(x;\theta^0) \in \Psi(\Omega)$. Therefore, to compare the learning speed of Algorithm~\ref{alg:API} and Algorithm~\ref{alg:RCTAC2}, 
both algorithms are implemented to find the nearly optimal policy and value function. We approximate the value function and policy using 3-layer fully-connected NNs. Each NN contains 2 hidden layers with 256 units per layer, activated by exponential linear units (ELUs). The outputs of the value and policy network are $V(x; \omega)$ and $\pi(x; \theta)$, using softplus unit and linear unit as activation functions, respectively. The training set consists of 256 states which are randomly chosen from the compact set $\Omega$ at each iteration. We set the stepsizes $\alpha_{\omega}$ and $\alpha_\theta$ to $0.01$ and employ  Adam for NN optimization.
\subsubsection{Result}
\begin{figure}[!htb]
\centering
\includegraphics[width=.98\linewidth]{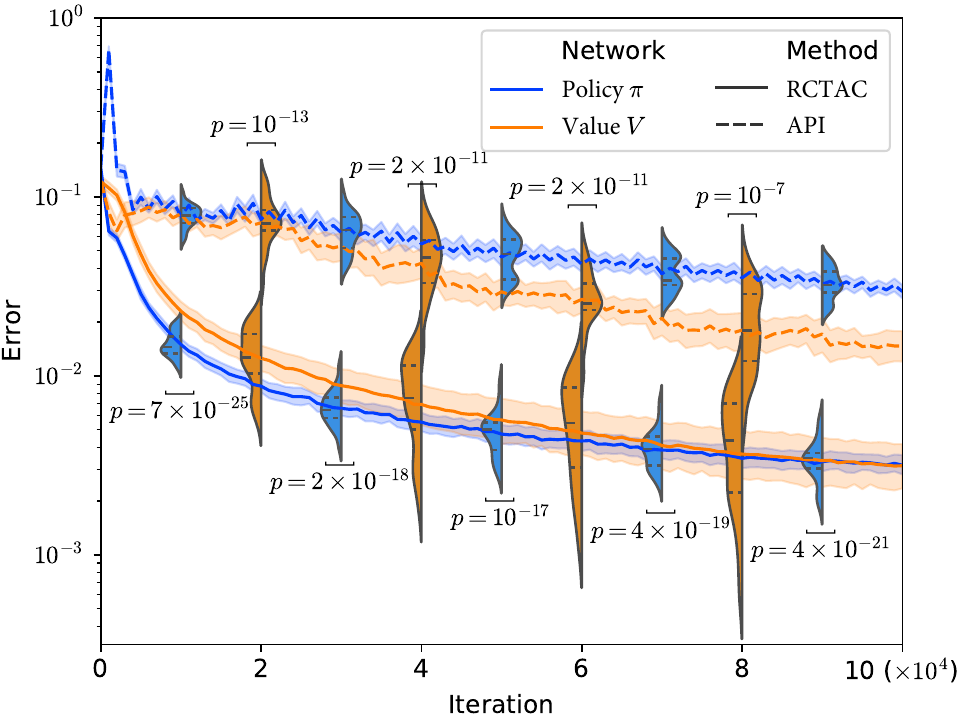}
\caption{RCTAC vs API performance comparison: Example I. Solid lines and shaded regions correspond to average values and $95\%$ confidence interval over 20 runs. One iteration corresponds to one NN update.}
\label{fig:example_1}
\end{figure}

Fig.~\ref{fig:example_1} displays the learning accuracy of the policy and value function at each iteration, measured by
\begin{equation}
\nonumber
\label{eq.e_pi_clculation}
e_{\pi} = \mathbb{E}_{x\in{X_0}}\left[ \frac{|\pi^\star(x)-\pi_{\theta}(x) |}{\max \limits_{x\in{X_0}}\pi^\star(x)-\min \limits_{x\in{X_0}}\pi^\star(x)}  \right], 
\end{equation}  
\begin{equation}
\nonumber
\label{eq.e_V_clculation}
e_V = \mathbb{E}_{x\in{X_0}}\left[ \frac{|V^\star(x)-V_{\omega}(x) |}{\max \limits_{x\in{X_0}}V^\star(x)-\min \limits_{x\in{X_0}}V^\star(x)}  \right],
\end{equation}  
where there are $500$ states in the test set ${X_0}$, which are randomly chosen from $\Omega$. We also draw violin plots in different iterations to show the precision distribution and 4-quartiles. 

It is clear from Fig.~\ref{fig:example_1} that both two algorithms can make the value and policy network approximation errors ($e_{\pi}$ and $e_V$) fall with iteration. And after $10^5$ iterations, both errors of Algorithm~\ref{alg:RCTAC2} are less than $0.4\%$. This indicates that Algorithm~\ref{alg:RCTAC2} has the ability to converge value function and policy to nearly optimal solutions. In addition, the t-test results in Fig. \ref{fig:example_1} show that both $e_{\pi}$ and $e_V$ of Algorithm~\ref{alg:RCTAC2} are significantly smaller than those of Algorithm~\ref{alg:API} ($p < 0.001$) under the same number of iterations. From the perspective of convergence speed, Algorithm~\ref{alg:RCTAC2} requires only about $10^4$ iterations to make both approximation errors less than 0.03, while Algorithm~\ref{alg:API} requires around $10^5$ steps. Based on this, Algorithm~\ref{alg:RCTAC2} is about 10 times faster than Algorithm~\ref{alg:API}. 

\begin{figure}
\centering
\captionsetup[subfigure]{justification=centering}
\subfloat[\label{subFig:example_1_state}]{\includegraphics[width=.98\linewidth]{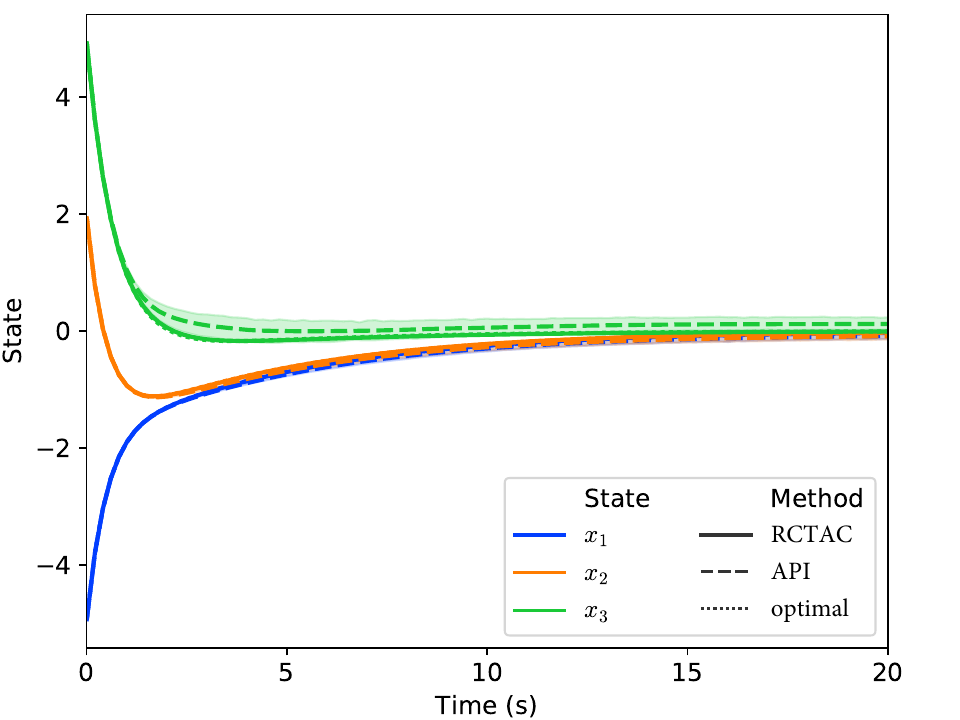}}\\
\subfloat[\label{subFig:example_1_u_V}]{\includegraphics[width=.98\linewidth]{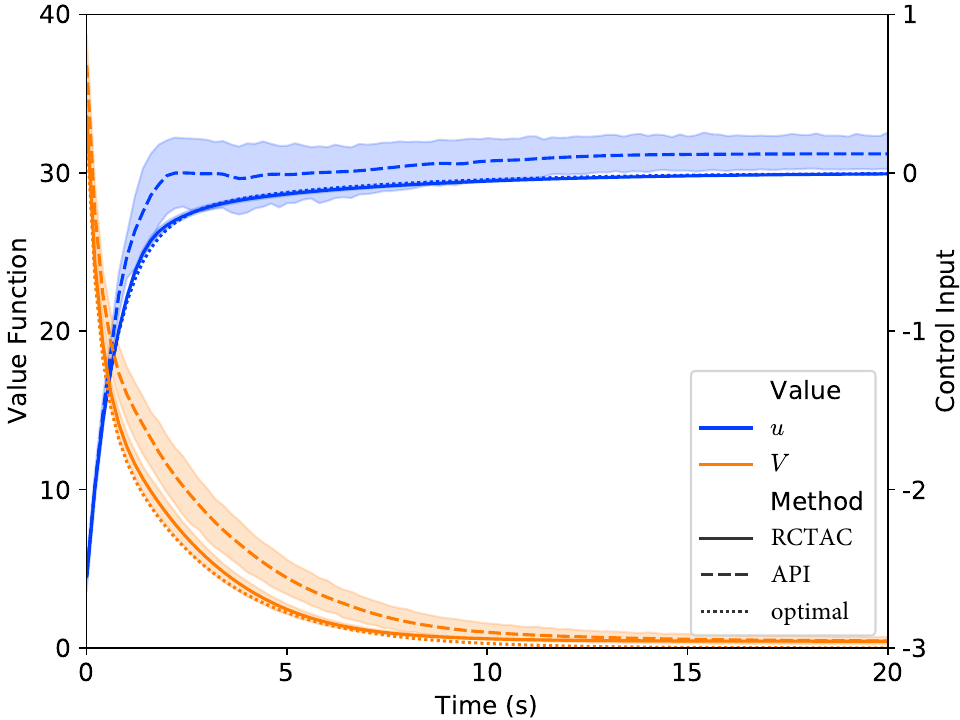}} 
\caption{Simulation results of different methods over 20 runs: Example I. (a) State trajectory. (b) Value function and control input. }
\label{fig:example_1_state}
\end{figure}

Fig. \ref{fig:example_1_state} shows the simulation results of the learned policy (after $10^5$ iterations) over 20 runs. It is obvious that policies learned by RCTAC perform much better than API. The curves of state, control input, and cost generated by RCTAC are almost consistent with the optimal solution. In summary, Algorithm \ref{alg:RCTAC2} is able to attain nearly optimal solutions and enjoys a significantly faster convergence speed compared to Algorithm \ref{alg:API}.

\subsection{Example II: Nonlinear and Input Non-Affine System}
\label{sec.Example II}
\subsubsection{Description}
Consider the vehicle path tracking task with nonlinear and input non-affine vehicle system derived in \cite{kong2015kinematic,li2017sharecontrol}. The desired velocity is 12 m/s and the desired tracking path is shown in Fig. \ref{fig:example_2_state}. Table \ref{tab.state} provides a detailed description of the states and inputs for this task, and Table \ref{tab.parameters} lists the vehicular parameters. The vehicle's movement is controlled by a bounded actuator, where  $a_x\in [-3, 3]~\mathrm{m/s}^2$ and $\delta \in [-0.35, 0.35]~\mathrm{rad}$. The dynamics of the vehicle is given by
\begin{equation*}
% \label{eq.model_vehicle}
x = 
\begin{bmatrix}
  v_y \\
  r \\
  v_x \\
  \phi \\
  y 
\end{bmatrix}
,u = 
\begin{bmatrix}
  \delta \\
  a_x 
\end{bmatrix}  
,f(x,u)=
\begin{bmatrix}
  \frac{F_{y\textnormal{r}}+F_{y\textnormal{f}}\cos\delta }{m} - v_x r\\
  \frac{- bF_{y\textnormal{r}}+aF_{y\textnormal{f}}\cos\delta }{I_{z}}\\
  v_y r+a_x - \frac{F_{y\textnormal{f}}\sin\delta}{m}\\
  r\\
 v_y \cos\phi + v_x \sin\phi 
\end{bmatrix}, 
\end{equation*} 
where $F_{y\ddagger}$ represents the lateral tire force and the subscript $\ddagger \in \{\textnormal{f,r}\}$ corresponds to the front or rear wheels.  It is calculated using the Fiala model:
\begin{equation}
\nonumber
\begin{split}
F_{y\ddagger}  =& -\mathrm{sgn}(\alpha_\ddagger)\min\Big\{  \left|\mu_\ddagger F_{z\ddagger}\right|,  \\
& \left|\tan(\alpha_\ddagger)C_\ddagger\Big(1+\frac{C_\ddagger^2(\tan\alpha_\ddagger)^2}{27(\mu_\ddagger F_{z\ddagger})^2} -  \frac{{C_\ddagger}\left |\tan\alpha_\ddagger \right |}{3\mu_\ddagger F_{z\ddagger}}  \Big)\right|\Big\}, 
\end {split}
\end{equation}
where $F_{z \ddagger}$ denotes the tire load, $\mu_\ddagger$ denotes the lateral friction coefficient, and $\alpha_\ddagger$ denotes the tire slip angle with
\begin{equation}
\nonumber
\alpha_\textnormal{f} = -\delta+\arctan (\frac{v_y+ar}{v_x}), \quad \alpha_\textnormal{r} = \arctan (\frac{v_y-br}{v_x}). 
\end{equation} 
The lateral friction coefficient is estimated by:
\begin{equation}
\nonumber
\mu_\ddagger = \frac{\sqrt{(\mu F_{z\ddagger})^2-F_{x\ddagger}^2}}{F_{z\ddagger}},
\end{equation} 
where $F_{x\ddagger}$ represents the longitudinal tire force, expressed as
\begin{equation}
\nonumber
F_{x\textnormal{f}} = \left\{
\begin{aligned}
  &0,  & a_x\ge0 \\
  &\frac{ma_x}{2}, & a_x<0
\end{aligned}
\right., \quad
F_{x\textnormal{r}} = \left\{
\begin{aligned}
  &ma_x,  & a_x\ge0 \\
  &\frac{ma_x}{2}, & a_x<0
\end{aligned}
\right..
\end{equation} 
The vertical loads on the front and rear wheels are approximated by 
\begin{equation}
\nonumber
F_{z\textnormal{f}} = \frac{b}{a+b}mg, \quad F_{z\textnormal{r}} = \frac{a}{a+b}mg.
\end{equation} 
The control objective is to minimize output tracking errors, which is formulated as
\begin{equation}
\nonumber
\begin{aligned}
\min_u \quad \int_{0}^{\infty} \Big\{0.4  (v_x -& 12)^2 + 80  y^2 + u^\top \left[ \begin{matrix}
    280  &  0  \\
    0  &  0.3  \\ 
 \end{matrix}\right]u\Big\} {\rm{d}} t
 \\
\rm{subject\ to} &\quad \dot{x} = f(x,u). 
\end{aligned}
\end{equation} 

\begin{table}[!htb]
\centering
\caption{List of state and input}
\label{tab.state}
\begin{tabular}{lll}
\toprule
Lateral  velocity &$v_y$ & [m/s]  \\
 Yaw rate at CG (center of gravity) &$r$ & [rad/s] \\
Heading angle between  trajectory \& vehicle &$\phi$ & [rad] \\
 Longitudinal velocity &$v_x$ & [m/s] \\
 Distance between trajectory
\& CG  &$y$ & [m] \\
\hline
Front wheel angle &$\delta$ & [rad]  \\
 Longitudinal acceleration &$a_x$ & [m/$\mathrm{s}^2$] \\
\bottomrule
\end{tabular}
\end{table}

\begin{table}[!htb]
\centering
\caption{Vehicle parameters}
\label{tab.parameters}
\begin{tabular}{lll}
\toprule
 Cornering stiffness at front wheel&$C_\textnormal{f}$ & -88000 [N/rad]  \\
 Cornering stiffness at rear wheel &$C_\textnormal{r}$ & -94000 [N/rad] \\
Mass &$m$ & 1500 [kg] \\
Distance from front axle to CG &$a$ & 1.14 [m] \\
Distance from rear axle to CG &$b$ & 1.40 [m] \\
Polar moment of inertia of CG &$I_z$ & 2420 [kg$\cdot\mathrm{m}^2$] \\
Tire-road friction coefficient&$\mu$ & 1.0 \\
\bottomrule
\end{tabular}
\end{table}
\subsubsection{Algorithm Details}
We employ 6-layer fully-connected NNs to represent $V_{\omega}$ and $\pi_{\theta}$, and the state input layer of each NN is followed by 5 fully-connected hidden layers, 32 units per layer. The activation function selection for the policy network is similar to that in Example I, with the exception that the output layer is activated by $\tanh(\cdot)$ with two units, multiplied by the vector $[0.35, 3]$ to accommodate the constrained control inputs. The training set consists of the current states of 256 parallel independent vehicles with different initial states, thereby obtaining a more realistic state distribution. We use Adam method to update two NNs, while the learning rates of value network and policy network are set to $8\times10^{-4}$ and $2\times10^{-4}$, respectively. Besides, we use $\eta=0.1$ to trade off the Hamiltonian term and the equilibrium term of the critic loss function (Remark \ref{remark.netarchiteture}).

\begin{figure}[!htb]
\centering
\includegraphics[width=.98\linewidth]{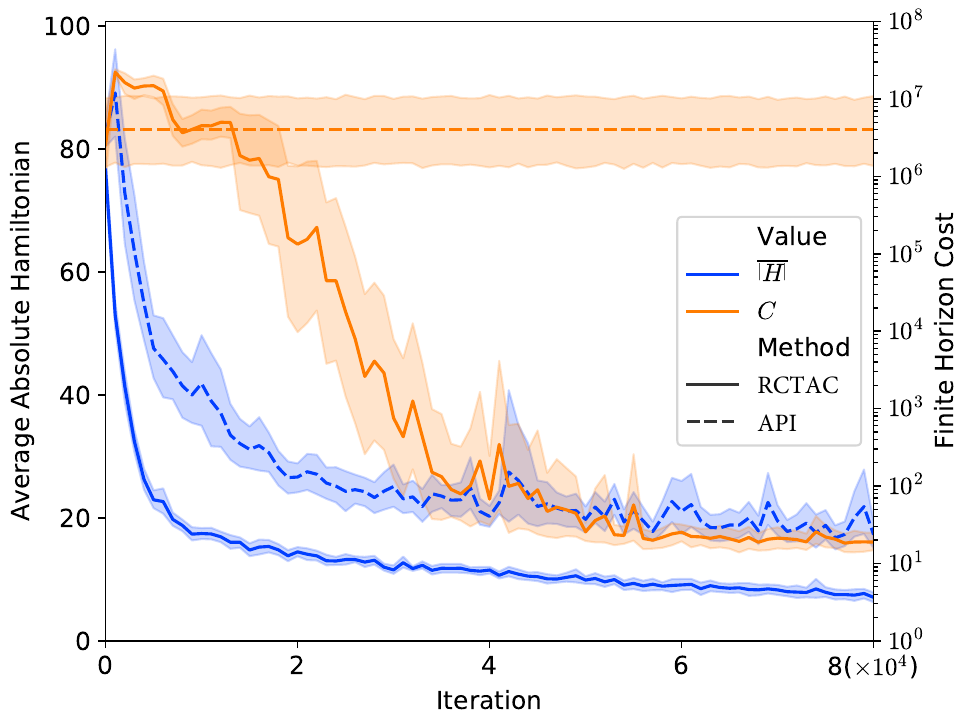}
\caption{RCTAC vs API performance comparison over 20 runs: Example II. }
\label{fig:example_2}
\end{figure}
\subsubsection{Result}
Fig. \ref{fig:example_2} shows the evolution of the average absolute Hamiltonian $\overline{\left | H \right |}$ of 256 random states and the training performance. The policy performance at each iteration is calculated by the cumulative running cost in 20s time domain
\begin{equation}   
\nonumber
\label{eq.finitecostfunction}
C = \int_{0}^{20} l(x(\tau),u(\tau)) {\rm{d}}\tau,
\end{equation}
where the initial state is randomly selected for each run. Since the initial policy is \emph{not} admissible, i.e., $\pi(x; \theta^0)\notin \Psi(\Omega)$, Algorithm~\ref{alg:API} can never make $\overline{\left | H \right |}$ close to 0, hence the terminal condition of policy evaluation can never be satisfied. Therefore, the finite horizon cost $C$ has no change during the entire learning process, i.e., Algorithm~\ref{alg:API} can never converge to an admissible policy if $\pi(x;\theta^0)\notin \Psi(\Omega)$. 

\begin{figure}[!htb]
\centering
\captionsetup[subfigure]{justification=centering}
\subfloat[\label{subFig:example2_control}]{\includegraphics[width=0.48\textwidth]{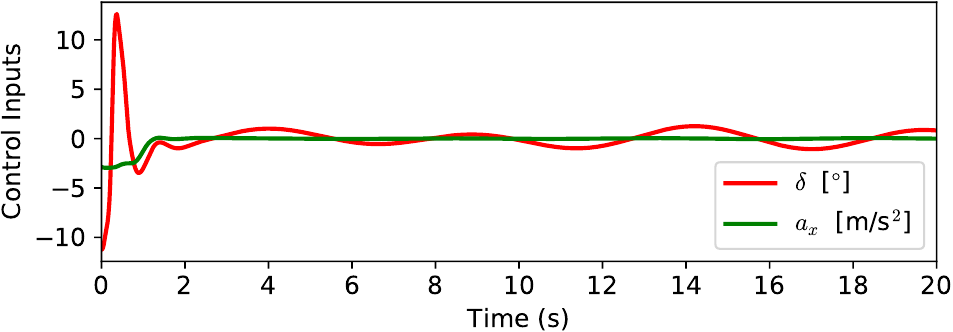}} \\
\subfloat[\label{subFig:example2_state}]{\includegraphics[width=0.48\textwidth]{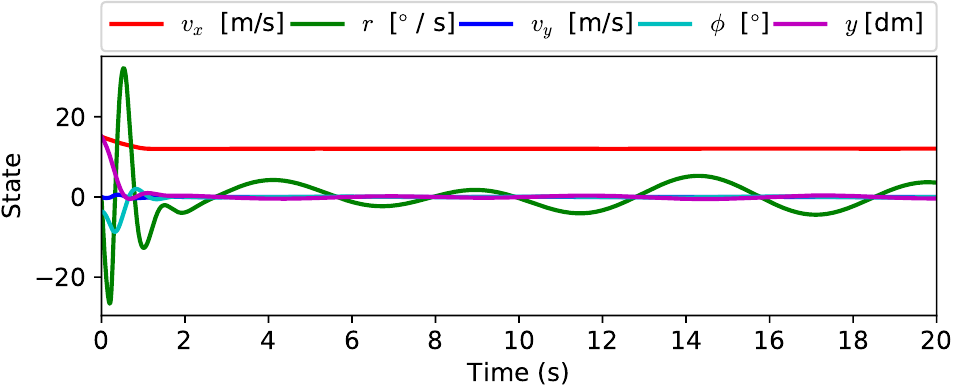}} \\
\subfloat[\label{subFig:sexample2_tra}]{\includegraphics[width=0.48\textwidth]{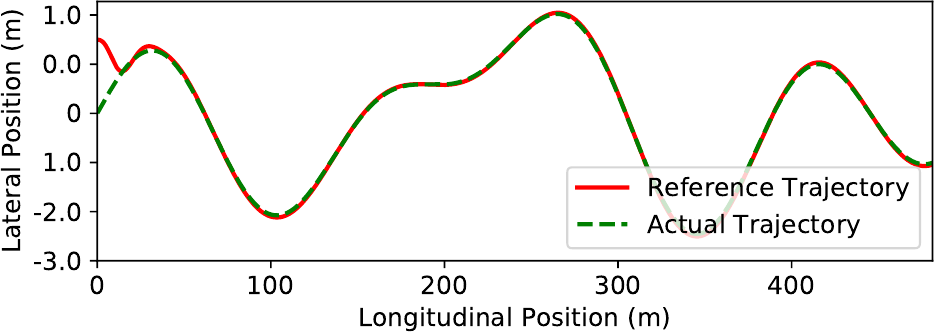}} \\
\caption{Simulation results: Example II. (a) Control inputs. (b) State trajectory. (c) Vehicle trajectory.}
\label{fig:example_2_state}
\end{figure}

On the other hand, $\overline{\left | H \right |}$ of Algorithm~\ref{alg:RCTAC2} can gradually converge to 0, while the finite horizon cost $C$ is also reduced to a small value during the learning process. Fig. \ref{fig:example_2_state} shows the control inputs and state trajectory controlled by the learned RCTAC policy. It can quickly guide the vehicle to the desired state, taking less than 0.5s for the case in Fig.~\ref{fig:example_2_state}. The results of Example II show that Algorithm~\ref{alg:RCTAC2} can solve the CT dynamic optimal control problem for general nonlinear and input non-affine CT systems with saturated actuators and handle inadmissible initial policies. 

In conclusion, these two examples demonstrate that the proposed RCTAC method is able to learn the nearly optimal policy and value function for general nonlinear and input non-affine CT systems without reliance on initial admissible policy. In addition, if the initial policy $\pi(x;\theta^0) \in \Psi(\Omega)$, the learning speed of Algorithm~\ref{alg:RCTAC2} is also faster than that of Algorithm~\ref{alg:API}.

\section{Experimental Validation}
\label{sec.experiment}
\begin{figure}[!htb]
\captionsetup{singlelinecheck = false,labelsep=period, font=small}
\centering{\includegraphics[width=0.99\linewidth]{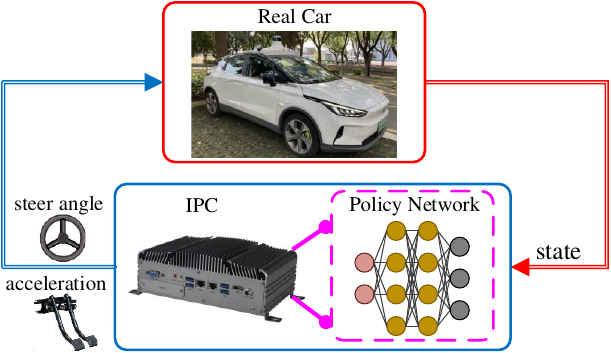}}
\caption{Real vehicle test pipeline.}
\label{f:architeture}
\end{figure}

In this section, we demonstrate the practical effectiveness of RCTAC by using the real-world path-tracking task~\cite{li2016intelligence,ge2022making}. The experiment pipeline is shown in Fig. \ref{f:architeture}. The test vehicle is GEELY Geometry C, which is equipped with a driving mode button, enabling it to switch between manual driving mode and automatic driving mode. We deploy the learned policy on the onboard industrial personal computer (IPC) with a 3.60 GHz Intel Core i7-7700 CPU. The deployed policy network is trained by the proposed RCTAC algorithm, where the expected velocity is set to 15 km/h, while other vehicle and algorithm parameters are consistent with Example II in Section \ref{sec.Example II}.  Provided with an ASENSING INS570D high-precision vehicle-mounted positioning system, the state of the vehicle can be accurately measured. After receiving the vehicle state information, the IPC records and calculates control instructions every 80ms, and sends the desired front wheel angle and longitudinal acceleration through the Controller Area Network (CAN), so as to realize the lateral and longitudinal control of the vehicle. 

\begin{figure}[!htb]
\centering
\captionsetup[subfigure]{justification=centering}
\subfloat[\label{subFig:experiment_control_input}]{\includegraphics[width=0.48\textwidth]{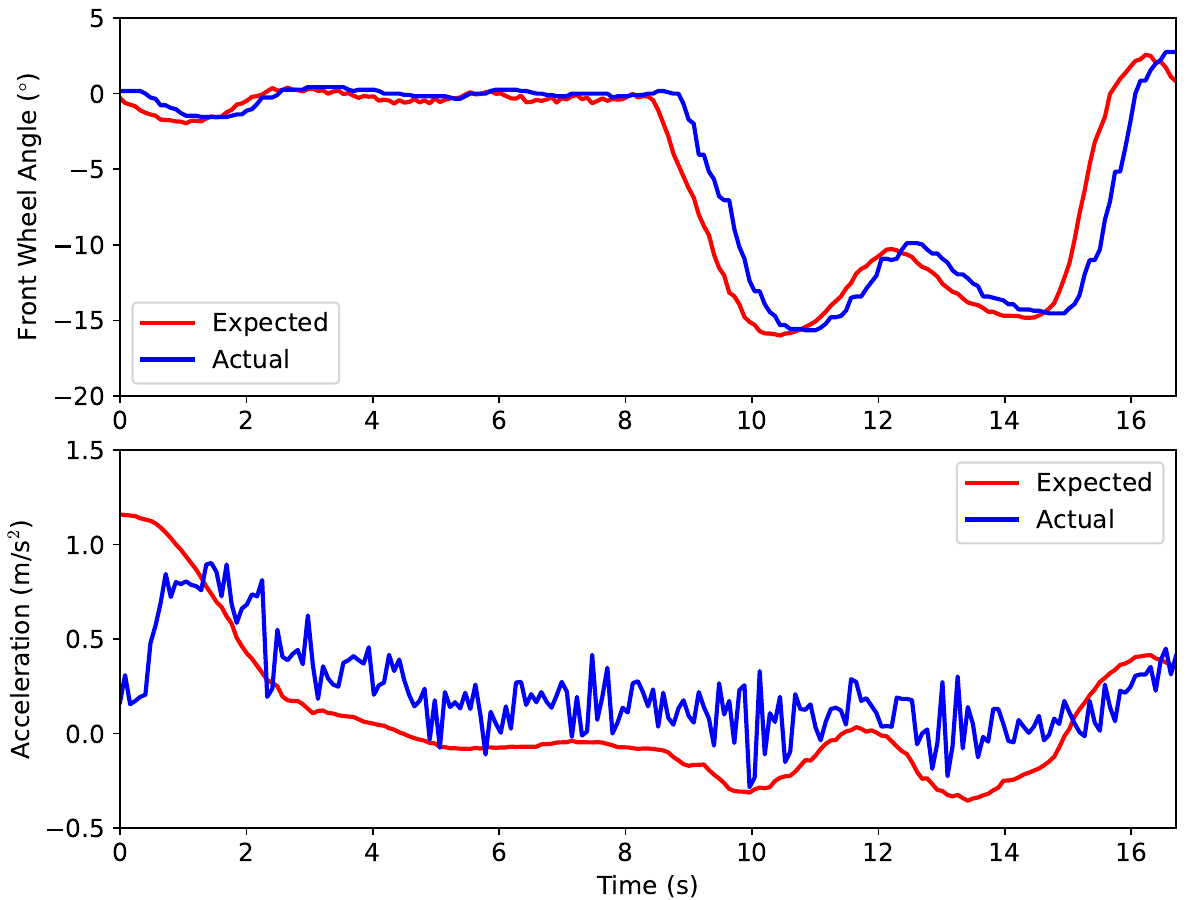}} \\
\subfloat[\label{subFig:experiment_state}]{\includegraphics[width=0.48\textwidth]{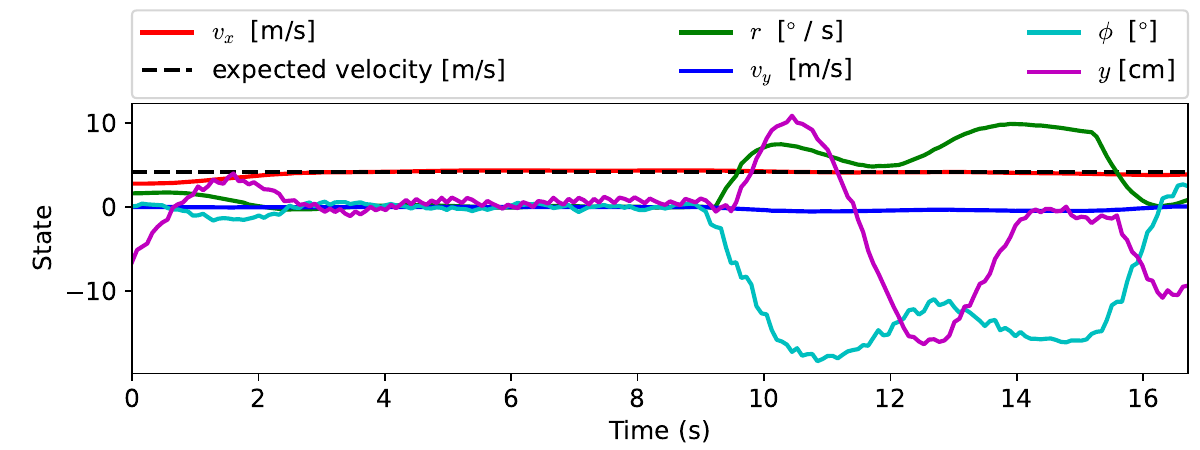}} \\
\subfloat[\label{subFig:experiment_global}]{\includegraphics[width=0.48\textwidth]{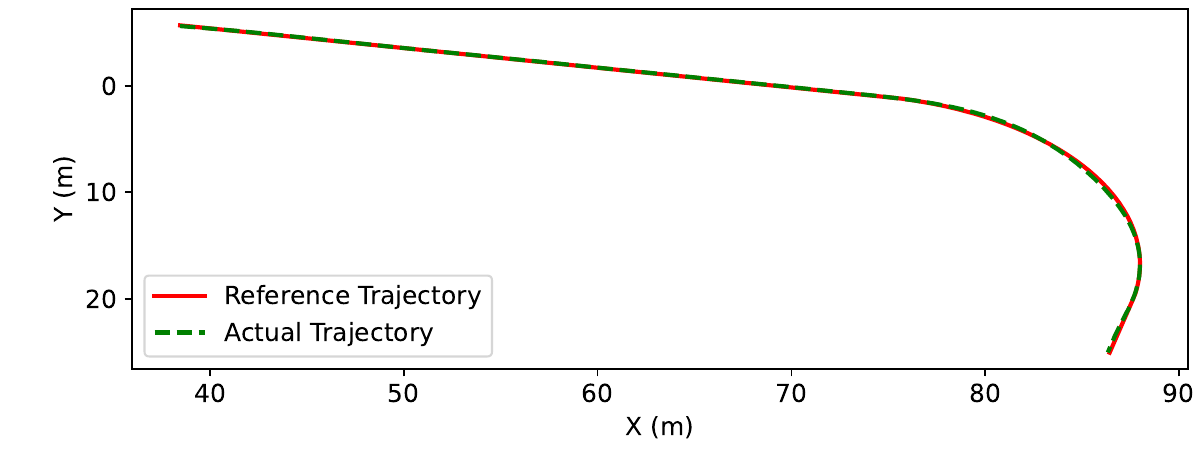}} \\
\caption{Experiment results: (a) Control inputs. The expected value corresponds to the policy output. (b) State trajectory. (c) Vehicle trajectory.}
\label{fig:experiment}
\end{figure}

The control inputs and state trajectory controlled by the deployed RCTAC policy are shown in Fig. \ref{fig:experiment}. From  Fig. \ref{subFig:experiment_control_input}, there exists a system response delay of about 0.5s between the expected and the actual control signal. Besides, the actual acceleration signals are noisy due to hardware limitations.  Despite these difficulties, our policy still makes the vehicle track the desired speed and trajectory smoothly and well (Fig. \ref{subFig:experiment_state},  \ref{subFig:experiment_global}). 

For comparison, model predictive control (MPC) with a prediction time domain of 20 steps is also deployed on the IPC to carry out practical experiments \cite{lin2021comparison}, where the well-known nonlinear programs solver IPOPT \cite{Andreas2006Biegler} is employed to solve the constructed MPC problem. The average results of 10 independent experiments are shown in Table~\ref{tab.compairson}. The root mean squared error (RMSE) of the heading angle between vehicle \& trajectory and that of the distance between CG \& trajectory 
\begin{equation}
    \nonumber
    \begin{aligned}
    \phi_{\rm RMSE} &= \sqrt{\frac{1}{m}{\sum\limits_{k = 1}^{m} {\phi}_{k}^{2}}}, \;
    y_{\rm RMSE} &= \sqrt{\frac{1}{m}{\sum\limits_{k = 1}^{m} y_{k}^{2}}}, 
    \end{aligned}
\end{equation}
are calculated to quantify the tracking performance. It can be found that RCTAC matches MPC in terms of tracking performance. Moreover, RCTAC shows great advantages in online decision-making efficiency, whose average single-step decision time is 91.17\% less than that of MPC.

\begin{table}[!htb]
\centering
\caption{Comparison Results}
\label{tab.compairson}
\tabcolsep = 0.08cm
\begin{tabular}{cccccc}
	\toprule
	      & $\phi_{\rm RMSE}$ [rad]  & $y_{\rm RMSE}$ [m]  & single-step decision time [ms] \\
	\midrule
	RCTAC & 0.0797  & 0.0568  & 3.67 \\
	MPC   & 0.0809 & 0.0601  & 41.57 \\
	\bottomrule
\end{tabular}
\end{table}

In conclusion, the vehicle experiment verifies the control effect of the proposed RCTAC algorithm in practical applications. It performs as well as MPC in our path-tracking task. Moreover, the way of offline training and online application makes the online calculation time of RCTAC much shorter than that of online optimization methods, such as MPC. This property is of significant importance for systems with limited computing resources.

\section{Conclusion}
\label{sec.conclu}
The paper proposed the relaxed continuous-time actor-critic (RCTAC) Algorithm~\ref{alg:RCTAC2}, along with the proof of convergence and optimality, for solving nearly optimal control problems of general nonlinear CT systems with known dynamics. The proposed algorithm can circumvent the requirements of ``admissibility'' and input-affine system dynamics (described in A1 and A2 of Introduction), quintessential to previously proposed counterpart algorithms. As a result, given an arbitrary initial policy, the RCTAC algorithm is shown to eventually converge to a nearly optimal policy, even for general nonlinear input non-affine system dynamics. The convergence and optimality are mathematically proven by using detailed Lyapunov analysis. We further demonstrate the efficacy and theoretical accuracy of our algorithm via two numerical examples, which yields a faster learning speed of the nearly optimal policy starting from an admissible initialization, as compared to the traditional approximate policy iteration (API) algorithm (Algorithm~\ref{alg:API}). In addition, a real-world path-tracking experiment is conducted to verify the practical performance of our method. Results show that compared with the MPC method, RCTAC has reduced the single-step decision time by 91.17\% without losing tracking performance. This indicates that our method has the potential to be applied in practical systems with limited computing resources. %In the future, we will combine other work about robust control and robust adaptive dynamic programming \cite{abu2008neurodynamic,liu2013robust} to extend the proposed algorithm to handle the uncertainties in the dynamic model.

\appendix
\section{Acknowledgment}
We would like to acknowledge Prof. Francesco Borrelli, Ms. Ziyu Lin, Dr. Yiwen Liao, Dr. Jiatong Xu, and Dr. Xiaojing Zhang for their valuable suggestions.

% \section*{Acknowledgment}
% We would like to acknowledge Dongjie Yu for his valuable suggestions.

\ifCLASSOPTIONcaptionsoff
  \newpage
\fi
\bibliographystyle{IEEEtran}
\bibliography{ref}

\end{document}